\documentclass[english,aps,pra]{revtex4-1}
\usepackage[T1]{fontenc}
\usepackage{amsmath,graphicx,amssymb,epsfig,babel,dsfont,color}

\begin{document}
	
\title{Conduction-radiation coupling between two\\distant solids interacting in near-field regime}

\author{Marta Reina}
\affiliation{Université Paris-Saclay, Institut d'Optique Graduate School, CNRS, Laboratoire Charles Fabry, 91127, Palaiseau, France}
\affiliation{ColibrITD, 91, rue du Faubourg Saint-Honoré 75008 Paris}
\author{Chams Gharib Ali Barura}
\affiliation{Université Paris-Saclay, Institut d'Optique Graduate School, CNRS, Laboratoire Charles Fabry, 91127, Palaiseau, France}
\author{Philippe Ben-Abdallah}
\affiliation{Université Paris-Saclay, Institut d'Optique Graduate School, CNRS, Laboratoire Charles Fabry, 91127, Palaiseau, France}
\author{Riccardo Messina}
\affiliation{Université Paris-Saclay, Institut d'Optique Graduate School, CNRS, Laboratoire Charles Fabry, 91127, Palaiseau, France}
\email{riccardo.messina@institutoptique.fr}

\date{\today}

\begin{abstract}
In the classical approach to deal with near-field radiative heat exchanges between two closely spaced bodies no coupling between the different heat carriers inside the materials and thermal photons is usually considered. Here we make an overview of the current state of studies on this coupling between solids of different sizes by paying a specific attention to the impact of the conduction regime inside the solids on conduction-radiation coupling. We also describe how the shape of solids affects this coupling. We show that this coupling can be at the origin of a drastic change of temperature profiles inside each body and of heat flux exchanged between them. These results could have important implications in the fields of nanoscale thermal management, near-field solid-state cooling and nanoscale energy conversion.
\end{abstract}

\maketitle

\section{Introduction}

Radiative heat transfer is the phenomenon through which two bodies at different temperatures can exchange energy even when separated by vacuum. A milestone in the study of this effect, dating back to the 19th century, is Stefan-Boltzmann's law, setting an upper bound for the flux two bodies at temperatures $T_1$ and $T_2$ can exchange: this upper limit, equal to $\sigma(T_1^4-T_2^4)$, $\sigma\simeq5.67\cdot10^{-8}\,\text{Wm}^{-2}\text{K}^{-1}$ being the Stefan-Boltzmann constant, can be realized only in the ideal scenario ot two blackbodies (i.e. bodies absorbing all the incoming radiation) exchanging heat. A second breakthrough in the study of radiative heat transfer was set much later, in the 1970s, by the development of fluctuational electrodynamics through the pioneering work of Rytov, Polder and van Hove~\cite{Rytov89,Polder71}. This theoretical framework describes each body as a collection of fluctuating dipoles whose statistical properties depend, by means of the fluctuation-dissipation theorem, on the temperature and optical properties of the body they belong to. This theory allowed to shine light on the very first experimental results~\cite{Hargreaves69} demonstrating the possibility to beat the blackbody limit in near-field regime, i.e. when their separation distance $d$ between the solids is small compared to the thermal wavelength $\lambda_\text{th}=\hbar c/k_\text{B}T$, of the order of 10$\,\mu$m at ambient temperature. More specifically, this is prone to happen when the two bodies support resonant modes of the electromagnetic field such as phonon-polaritons (for polar materials) and plasmons (for metals)~\cite{Joulain05} or even a continuum of evanescent modes such as hyperbolic modes~\cite{Biehs12}. 

The unveiling of a near-field flux amplification paved the way to numerous experiments (see \cite{Song15,Cuevas18,Biehs21} for some review papers) in a variety of geometries (including e.g. plane--plane, sphere--plane and tip--plane) and for several different materials. Parallel to the experimental investigations, several ideas of applications have been put forward, ranging from energy-conversion devices~\cite{DiMatteo01,Narayanaswamy03,Basu07,Fiorino18} to heat-assisted data recording~\cite{Challener09,Stipe20}, infrared spectroscopy~\cite{DeWilde06,Jones12}, and thermotronics~\cite{BenAbdallah13,BenAbdallah15}, namely the conception of thermal equivalents of electrical circuit elements.

Although the vast majority of experiments have confirmed the theoretical predictions, some of them have observed deviations from it, both in the extreme near-field scenario~\cite{Kloppstech17} (nanometer and sub-nanometer range of distances) and at tens of nanometers~\cite{Kittel05,Shen09}. More specifically, some experiments~\cite{Kloppstech17} observed an amplification of the flux, whereas others~\cite{Kittel05,Shen09} highlighted a saturation effect. The explanation of such inconsistencies between theory and experiment, to date unsolved, has stimulated the theoretical investigations in several directions. First, is has been suggested that non-local effects must be taken into account in order to describe the energy exchange between metals at very short separation distances~\cite{Ford84,Chapuis08}. Moreover, in the extreme near field the participation of other heat carriers (phonons and electrons) could affect significantly the exchanged flux ~\cite{Zhang18,Alkurdi20,Volokitin19,Volokitin20,Volokitin21,Tokunaga21,Tokunaga22,Guo22,GomezViloria23}, but these can only play a role below a few nanometers. Finally, some works have also explored in this sense the transition between conduction and radiation~\cite{Joulain08,Chiloyan15}.

\begin{figure}
	\center\includegraphics[width=9cm]{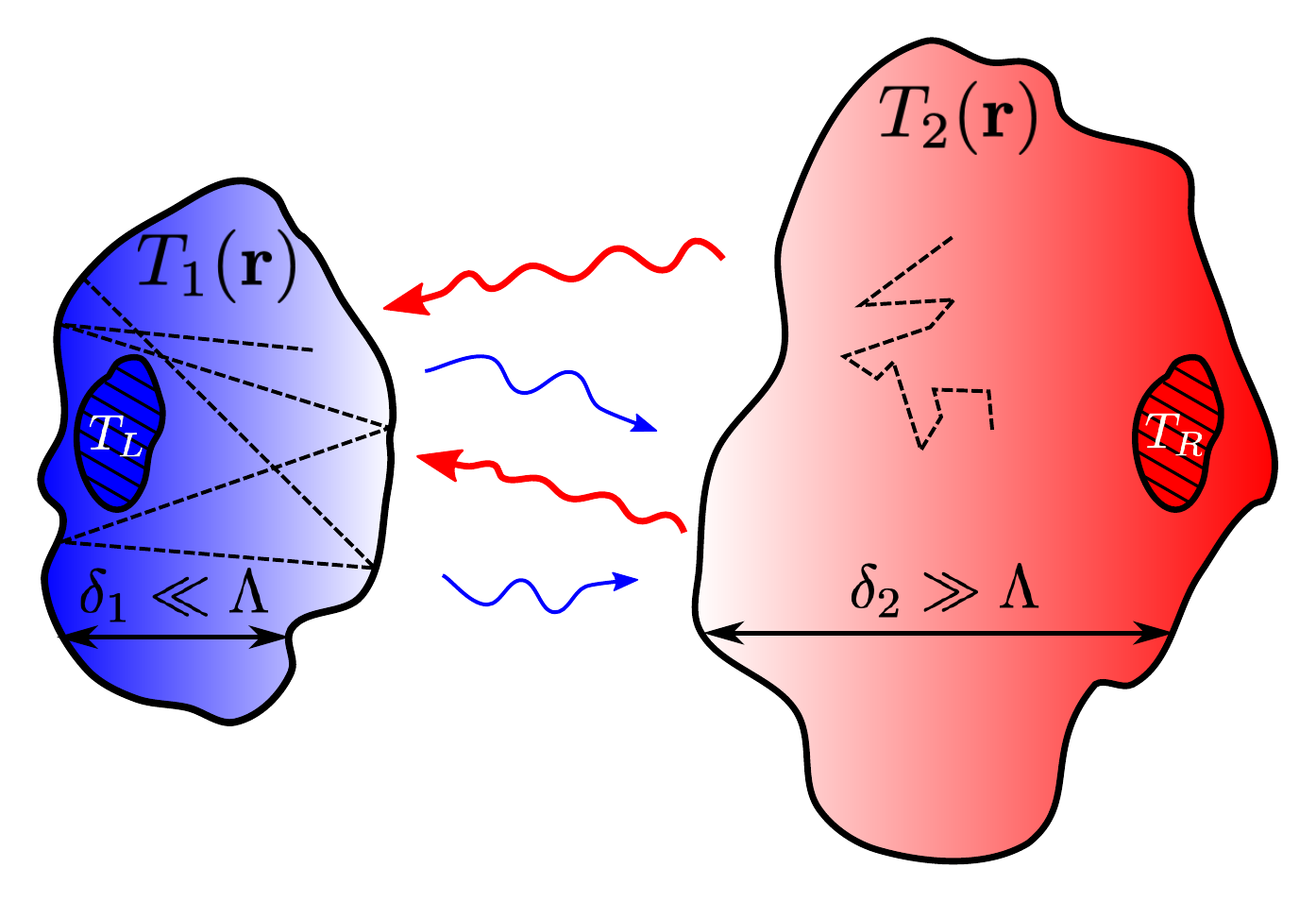}
	\caption{Configuration involging two arbitrarily-shaped bodies of finite size and kept at different temperatures $T_L$ and $T_R$ by two thermostats. The two bodies exchange heat radiatively, while conduction takes place inside each of them. The size $\delta$ of each body compared to the phonon mean free path $\Lambda$ dictates the conductive transport regime. For $\delta\ll\Lambda$ (left) the heat transport is ballistic (no collision events during phonon trajectories), while for $\delta\gg\Lambda$ (right) the heat transport is diffusive (many collision events). The coupling of conductive and radiative heat transfer incudes, in general, two temperature profiles $T_{1,2}(\mathbf{r})$ inside the two bodies. Reproduced from \cite{Reina2021a}.\label{fig_geometry}}
\end{figure}

A further effect which could be at the origin of a deviation with respect to the predictions of fluctuational electrodynamics is the coupling between conduction acting inside each body and near-field radiative heat transfer between them. In order to get an idea of the possible impact of this effect, we can visualize the typical theoretical system, as shown in Fig.~\ref{fig_geometry}. Two bodies are kept at temperatures $T_L$ and $T_R$ by two thermostats, locally connected to them. In almost all theoretical works on near-field radiative heat transfer, it is assumed that conduction inside each body is so efficient (compared to the energy exchange mediated by radiation) that the temperature can be assumed to be uniform in each body and equal to the one imposed by the thermostat. This allows to define properly the radiative heat transfer between two bodies at two given temperatures $T_L$ and $T_R$. Nevertheless, the strong dependence of near-field radiative heat transfer on the materials involved and, more importantly, on the separation distance suggests that the two effects could compete in some ranges of parameters. This would imply the existence of a temperature profile within each body (as depicted in Fig.~\ref{fig_geometry}) and then, in turn, a modification of the flux exchanged through radiation.

During the last years, we have performed a comprehensive study of the impact of this coupling~\cite{Messina2016a,Messina2016b,Jin17a,Jin17b,Reina2021a,Reina2021b,GharibAliBarura22}, which is the topic of this review paper. More specifically, we have first studied this effect in the simple geometry of two parallel slabs and in the diffusive regime, as discussed in Sec.~\ref{sec_slabs}. Following this first analyis, we have investigated the role played by the size of the two bodies. As a matter of fact, as shown pictorially in Fig.~\ref{fig_geometry}, this can have an impact on the conduction tranport regime inside each body. These results are discussed in Sec.~\ref{sec_Boltzmann}. Finally, in order to account for the variety of geometries employed in experiments, we have also studied the same coupling effect in different geometries, as discussed in Sec.~\ref{sec_geometry}. Finally, some conclusions are given in Sec.~\ref{sec_conclusions}.

\section{Slab-slab configuration in the diffusive conduction regime}\label{sec_slabs}

\begin{figure}
		\centering
		\includegraphics[width=9cm]{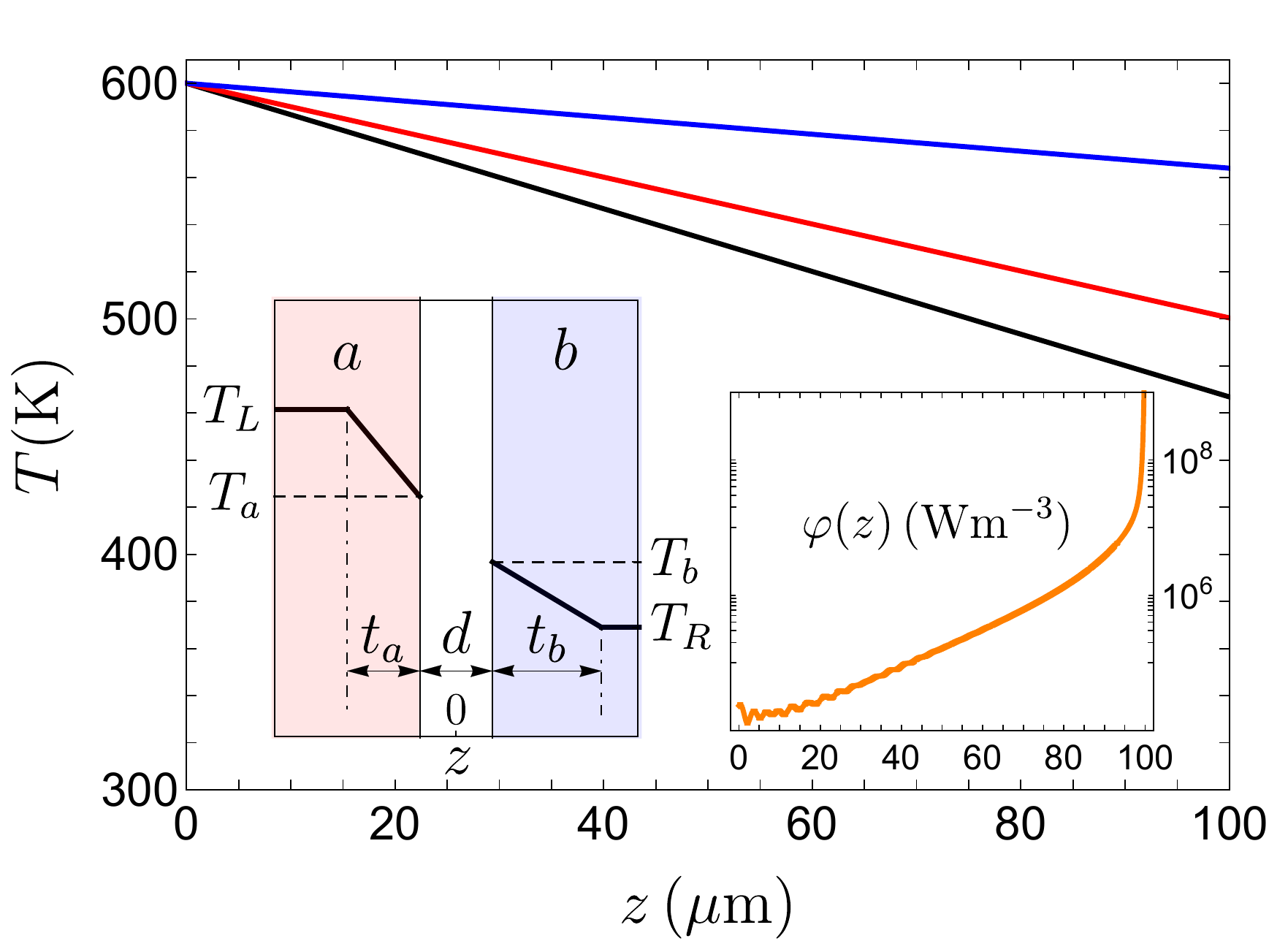}
	\caption{Geometry of two parallel slabs separated by a $d$-thick vaxcuum gap. In the left (right) slab the temperature can vary with respect to $T_L$ ($T_R$) over a thickness $t_a$ ($t_b$). Temperature profile along the left slab for two silica slabs with $t_a = t_b = 100\,\mu$m and $(T_L,T_R) = (600, 300)\,$K. The lines correspond to $d = 10\,$nm (black), 20\,nm (red) and 50\,nm (blue). The right inset shows the position-dependent radiative flux $\phi(z)$ for $d = 100\,$nm. Reproduced from \cite{Messina2016a}.\label{fig_slabs1}}
\end{figure}

The simplest geometry to study the effect of conduction-radiation coupling is indeed the one involving two parallel finite-thickness slabs separated by a vacuum gap of thickness $d$, as represented in the inset of Fig.~\ref{fig_slabs1}. To describe the action of two thermostats connected to the two bodies, we assume that the temperature in the first (second) body is fixed at $T_L$ ($T_R$) except over a region of thickness $t_a$ ($t_b$).

In order to further simplify the problem, we also assume that the thickness of the two slabs is large enough to safely treat the conduction problem in the Fourier diffusive regime. In this case, the coupled equation to be solved reads
\begin{equation}
	\frac{\partial}{\partial z} \left[ \kappa(z)\frac{\partial}{\partial
		z} T(z)\right]+\int \mathrm{d}z'\,\varphi(z',z)=0.
	\label{eq:heat}
\end{equation}
In this equation $\kappa(z)$ is the bulk Fourier conductivity at point $z$, respectively, whereas $\varphi(z',z)$ representes the radiative power per unit volume emitted at a point $z'$ and absorbed at a point $z$. At this stage, the expression of the radiative term $\varphi(z',z)$ is needed. This energy exchange can be, for example, calculated by means of a framework introduced to calculate both Casimir forces and radiative heat transfer and based on the knowledge of the scattering operators of the bodies involved~\cite{Messina11a,Messina11b,Messina14,Latella17}. In order to account for the temperature profiles, we assume that each body is divided in slabs of infinitesimal thickness, of which the scattering coefficients are known analytically, and apply the scattering approach to deduce the radiative heat transfer. Limiting ourselves to the contribution stemming from evanescent waves in transverse magnetic polarization (dominating in the near field between polar materials~\cite{Joulain05}) we write the flux as the frequency and wavevector integral $\varphi(z_a,z_b)=\int_0^{\infty}\mathrm{d}\omega\int_{\omega/c}^{\infty}\mathrm{d}\beta~\varphi_a(\omega,\beta;z_a,z_b)$, where the spectral flux can be expressed as (see \cite{Messina2016a} for more details):
\begin{equation}\label{eq:slab}
		\varphi(\omega,\beta;z_a,z_b)=\frac{4\beta}{\pi^2}(r''k_{zm}'')^2\frac{e^{-2k_z''d}\,e^{-2k_{zm}''(z_b-d/2)}}{|1-r^2e^{-2k_z''d}|^2}\Bigl(N[\omega,T(z_a)]-N[\omega,T(z_b)]\Bigr),
\end{equation}
where $\beta$ is the parallel ($x$--$y$) component of the wavevector, while
$k_z=\sqrt{\omega^2/c^2-\beta^2}$ and
$k_{zm}=\sqrt{\varepsilon\omega^2/c^2-\beta^2}$ are the perpendicular
components in vacuum and inside the slabs, respectively. We also introduced the Fresnel reflection coefficient of a slab, given by $r=(\varepsilon k_z-k_{zm})/(\varepsilon k_z+k_{zm})$. Finally, in Eq.~\eqref{eq:slab} $a''$ represents the imaginary part of $a$ and
and
\begin{equation}
	N(\omega,T) = \biggl[\exp\biggl(\frac{\hbar\omega}{k_B T}\biggr) - 1\biggr]^{-1}.
\end{equation}

The coupled heat equation \eqref{eq:heat}, combined with the flux expression \eqref{eq:slab}, can be solved numerically. Nevertheless, two further approximations can be performed and allow us to obtain an analytical expression for both the temperature gap $T_a-T_b$ between the two slabs (across the vacuum gap, see Fig.~\ref{fig_slabs1}) and the exchanged flux. We can first assume that the radiative energy exchange takes places over a tiny thickness close to the vacuum interface of each body, allowing us to treat it as a surface term, i.e. as a boundary condition. Moreover, inspired by the results of fluctuational electrodynamics, we can assume that the total flux exchanged radiatively can be expressed as $\phi\simeq h_0(T_a-T_b)/d^2$. In this simplified expression, the flux depends only on the two temperatures at the interfaces and shows the known $d^{-2}$ divergence. These approximations, whose validity has been verified numerically, lead to the following analytical solutions
\begin{equation}
	\label{analytics}
	\frac{T_a-T_b}{T_L-T_R}=\left(1+\frac{2th_0}{\kappa
		d^2}\right)^{-1},\quad\frac{\varphi}{T_L-T_R}=\frac{h_0}{d^2} \left(\frac{T_a-T_b}{T_L-T_R}\right).
\end{equation}

\begin{figure}
		\centering
		\includegraphics[width=9cm]{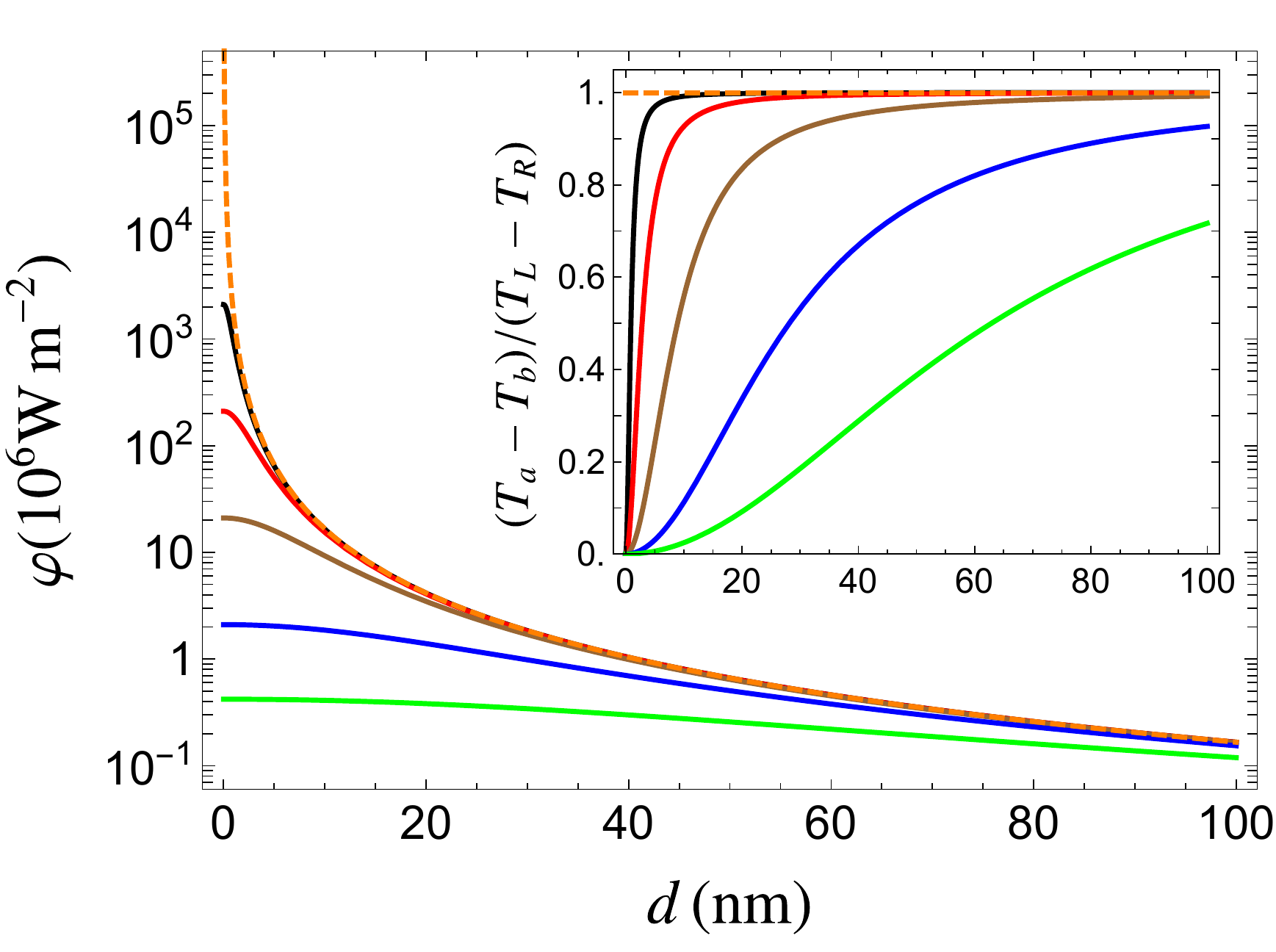}
		\caption{Flux $\varphi$ and temperature difference $T_a-T_b$ across the vacuum gap (inset) as a function of $d$ between two silica slabs with $T_L=600\,$K and $T_R=300\,$K. The solid lines correspond to different thicknesses: 100\,nm (black), $1\,\mu$m (red), $10\,\mu$m (brown), $100\,\mu$m (blue) and $500\,\mu$m (green). The orange dashed line corresponds to the absence of temperature gradients. Reproduced from \cite{Messina2016a}.\label{fig_slabs2}}
\end{figure}

The impact of conduction-radiation coupling is shown quantitatively in Figs.~\ref{fig_slabs1} and \ref{fig_slabs2}. Figure \ref{fig_slabs1} concerns two silica slabs having $t_a=t_b=100\,\mu$m and $(T_L,T_R)=(600,300)\,$K. The temperature profile is shown inside the left slab for three different distances: it shows that the temperature can decrease by more than 100\,K in the left slab for the smallest distance considered. Moreover, the inset of Fig.~\ref{fig_slabs1}, showing the numerically-calculated distribution of flux absorbed inside the slab, shows that it is highly peaked around the vacuum interface, as assumed.

Figure \ref{fig_slabs2} shows the temperature difference across the gap $T_a-T_b$ (normalized with respect to $T_L-T_R$, inset) and the exchanged flux (main part) for different slab thicknesses (see caption of Fig.~\ref{fig_slabs2}). We clearly observe that not only is a tempetaure profile inuced by the coupling, with an effect growing with the slab thickness, but that also the flux is strongly modified with respect to the scenario of absence of coupling (orange dashed line in Fig.~\ref{fig_slabs2}). The flux tends to saturate for $d$ going to zero, and both the saturation value and the characteristic distance at which the distance-dependent flux deviates from the no-coupling scenario depend strongly on the thickness.

\begin{figure}
		\centering
		\includegraphics[width=9cm]{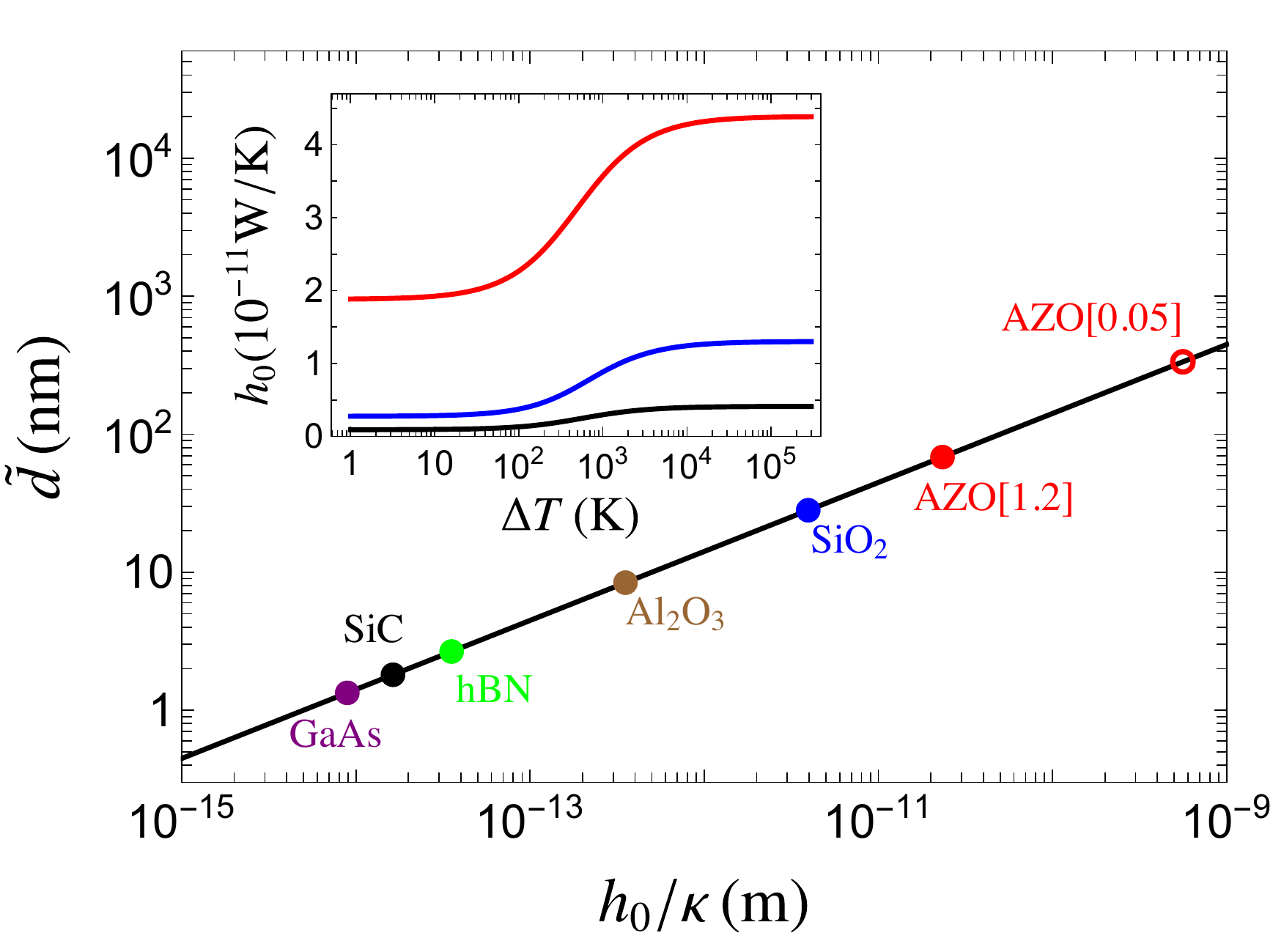}
	\caption{Characteristic distance $\tilde{d}$ of 
		conduction--radiation coupling (see text) for $t_a=t_b=100\,\mu$m and
		different materials. AZO[1.2] and
		AZO[0.05] denote aluminum zinc oxides of conductivities
		$\kappa=1.2\,$W/m$\cdot$K and $\kappa=0.05\,$W/m$\cdot$K,
		respectively (see \cite{Messina2016a} for more details). The inset shows $h_0$ as a function of $\Delta T=T_L-300\,$K, for AZO (red), silica (blue) and SiC (black). Reproduced from \cite{Messina2016a}.\label{fig_slabs3}}
\end{figure}

In order to make this more quantitative it is interesting to define a characteristic coupling distance $\tilde{d}$, such that at this distance the temperature gradient across the vacuum gap equals half of $T_L-T_R$. At this distance we have
$T_a-T_b=\frac{1}{2}(T_L-T_R)$ and $\varphi = \frac{1}{2}
h_0(T_L-T_R)/\tilde{d}^2$. This distance $\tilde{d} = \sqrt{2th_0/\kappa}$ depends both on the thickness and on the material-dependent $h_0/\kappa$ parameter, quantifying the competition between radiative exchange (through the conductance $h_0$) and conductive transport (through the conductivity $\kappa$). Figure \ref{fig_slabs3} shows $\tilde{d}$ as a function of this ratio for $t=100\,\mu$m. On top of this curve we highlight some examples of materials, showing that this characteristic distance can vary from some to tens or hundreds of nanometers, making the experimental observation of conduction--radiation coupling in principle feasible for some materials and thicknesses. We conclude this section mentioning that a previous study was performed in this sense~\cite{Wong14}, but applied only to a specific configuration and not exploring the strong dependence on the choice of materials and thicknesses.

\section{The impact of slab thickness: from the diffusive to the ballistic regime}\label{sec_Boltzmann}

The results presented in the previous section are based on the assumption that the thickness of the two bodies exchanging heat is large compared to the mean free path of phonons inside them, in the micron range for typical polar materials. Nevertheless, because of the peculiar transport regimes which can arise for conduction depending on the thickness, we have also extended our study to the more general configuration of arbitrary thickness. Boltzmann's transport equation is the mathematical tool allowing us to fully grasp the transition between the transport regimes, and more specifically between the two extreme ones, namely ballistic and diffusion regimes.

At a given frequency $\omega$ (not explicitly shown) and in the relaxation time approximation, this equation reads
\begin{equation}
	\frac{\partial f_p(t, \omega,\mathbf{r},\Omega)}{\partial t}+\mathbf{v}_{g,p}(\omega)\cdot\nabla f_p(t, \omega,\mathbf{r},\Omega)=-\frac{f_p(t, \omega,\mathbf{r},\Omega)-f_0(\omega)}{\tau_p(\omega,T(\mathbf{r}))}.
	\label{Eq:Boltzmann_Eq}
\end{equation}
The unknown of this equation is the distribution function $f$ associated to the heat carriers within the solid for each polarization $p$, at time $t$, frequency $\omega$, solid angle $\Omega$ and position $\mathbf{r}$. Moreover, $\mathbf{v}_{g,p}(\omega)=\nabla_\mathbf{k}\omega_p$ is the group velocity of carriers at polarization $p$ and frequency $\omega$, $f_0$ is the equilibrium distribution (Fermi-Dirac for electrons and Bose-Einstein for phonons) and $\tau_p$ the heat-carrier relaxation time.

In order to be solved, this equation has to be coupled to the one governing the time evolution of the internal energy density $u$, which reads
\begin{equation}
	\frac{\partial u(\mathbf{r},t)}{\partial t}=P_{\text{rad}}(\mathbf{r},t)+P_{\text{cond}}(\mathbf{r},t),
	\label{Eq:energy_Eq}
\end{equation}
$P_{\text{rad}}$ denoting the radiative power locally dissipated per unit volume within a given body and coming from the other one, which can be calculated within a fluctuational-electrodynamics approach (see \cite{Reina2021a} for more details). On the other hand, $P_{\text{cond}}$ denotes the conductive power per unit volume at position $\mathbf{r}$, which is connected to the distribution function $f$ through the relation
\begin{equation}
	\varphi_{\text{cond}}(t,\mathbf{r})=\sum_p\int_{4\pi} d\Omega\int d\omega\,\hbar \omega\, \mathbf{v}_{g,p}(\omega) f_p(t, \omega,\mathbf{r},\Omega)\frac{D_p(\omega)}{4\pi}, 
	\label{Eq:flux_cond}
\end{equation}
where $D_p(\omega)$ represents the density of states.

We have solved the two coupled equations in the simple geometry of two parallel SiC slabs: the left one, denoted by index 1, is connected to a thermostat at temperature $T_L=400\,$K, whereas slab 2 (on the right) is connected to one at $T_R=300\,$K. Concerning the boundary conditions, two different cases must be taken into account: for the edges in contact with vacuum phonons hitting the surface are scattered specularly (specular reflection), whereas phonons colliding against the thermostat are scattered in all directions (diffuse reflection)~\cite{ReinaPhD}. Exploiting Boltzmann's equation, we are allowed to let the slab thickness vary in a wide range of values, from the ballistic to the diffusive regime.

\begin{figure}
		\centering
		\includegraphics[width=7cm]{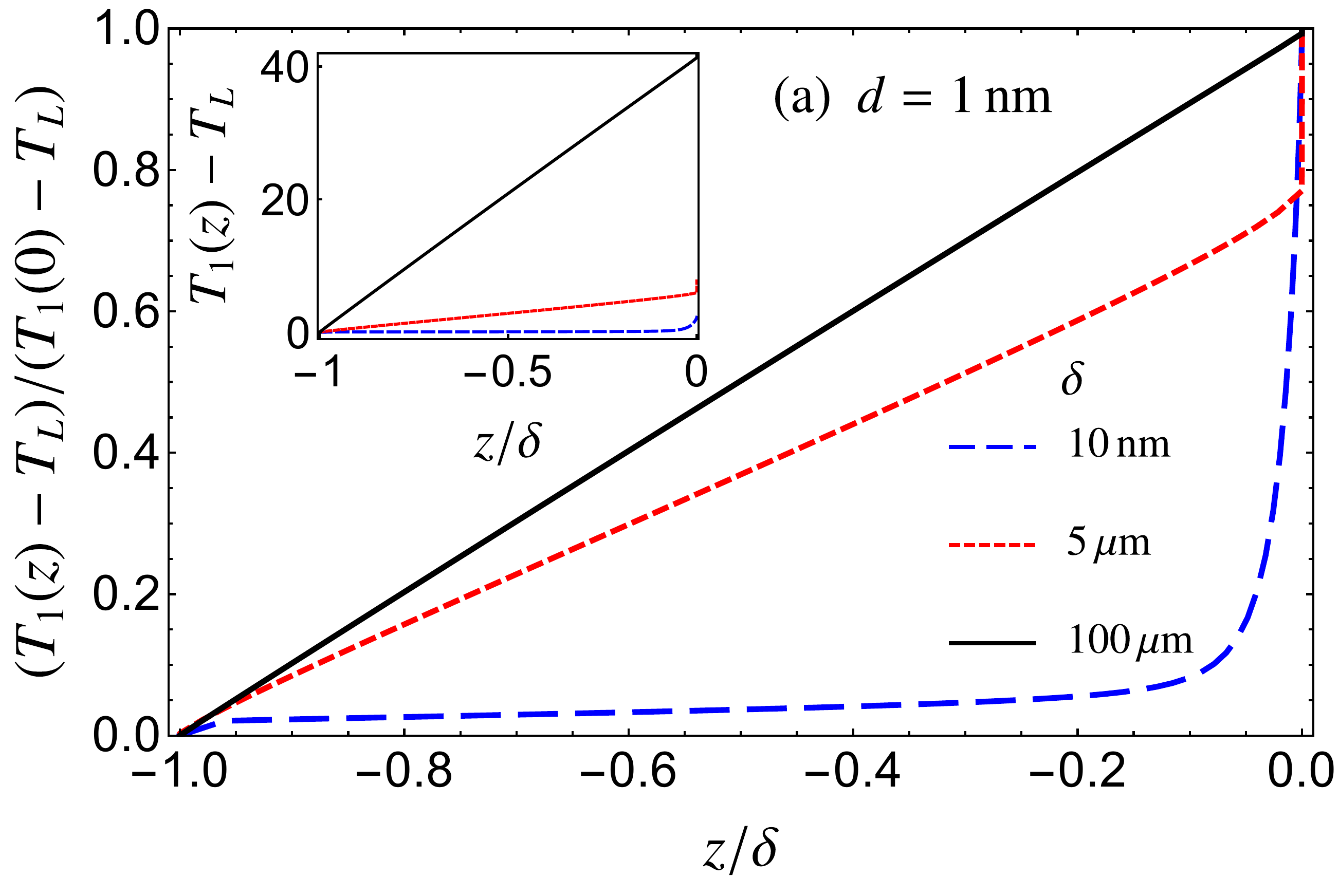}\hspace{1.5cm}
		\includegraphics[width=7cm]{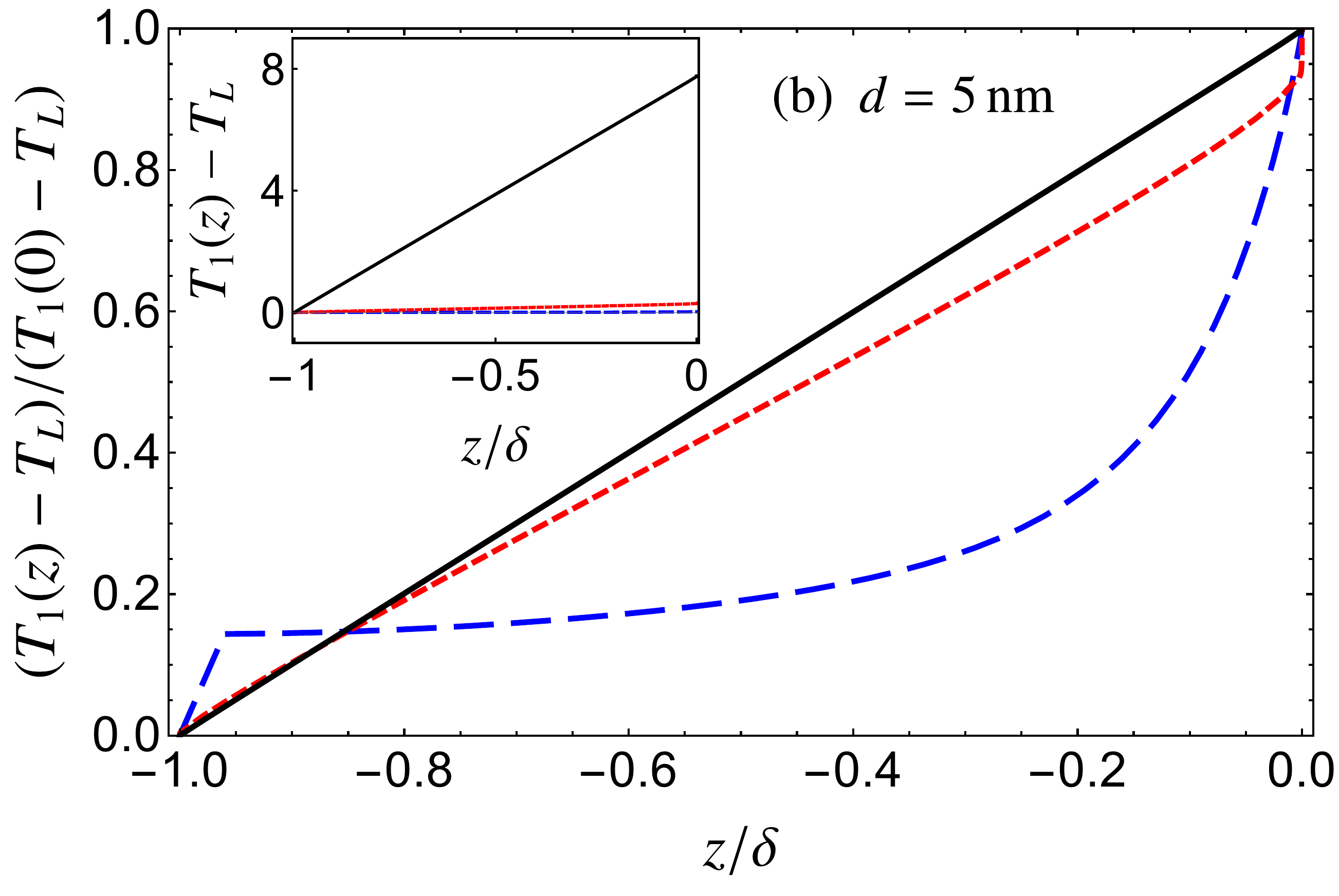}
	\caption{(a) Steady-state temperature (inset) and normalized temperature profile inside the left slab for different thicknesses and a separation distance $d=1\,$nm. (b) Same as (a) for $d=5\,$nm. Reproduced from \cite{Reina2021a}.\label{fig_Boltzmann1}}
\end{figure}

The results at two separation distances ($d=1$ and $5\,$nm) are shown in Fig.~\ref{fig_Boltzmann1}. The main part of the two plots shows the temperature profile inside slab 1, normalized to the temperature difference across it $T_1(0)-T_L$. This allows to highlight a signature of the conductive tranport regime in the shape of the temperature profile. More specifically, while for the largest thickness considered $T_1(z)$ becomes almost linear (as it should be in the strictly diffusive regime, according to Fourier law), when decreasing the thickness we observe a transition towards a significantly different behavior. In this ballistic-like scenario the temperature profile tends to a uniform distribution, excluding the region close to the thermostat ($z\simeq-\delta$), where $T_1(z)$ is almost discontinuous (Casimir regime), and close to the vacuum gap, where $T_1(z)$ shows a steep increase physically connected to the fact that most to the radiative flux is absorbed close to the boundary.

While this discussion highlights the impact of transport regime on the shape of the temperature profile, the information regarding the quantitative impact of distance and thickness is contained in the insets of Fig.~\ref{fig_Boltzmann1}. It is clear that an observable temperature profile (up to tens of degrees) can indeed arise, but mainly for large thicknesses (tens of microns) and small distances (below 5\,nm). 

\begin{figure}
		\centering
		\includegraphics[width=7cm]{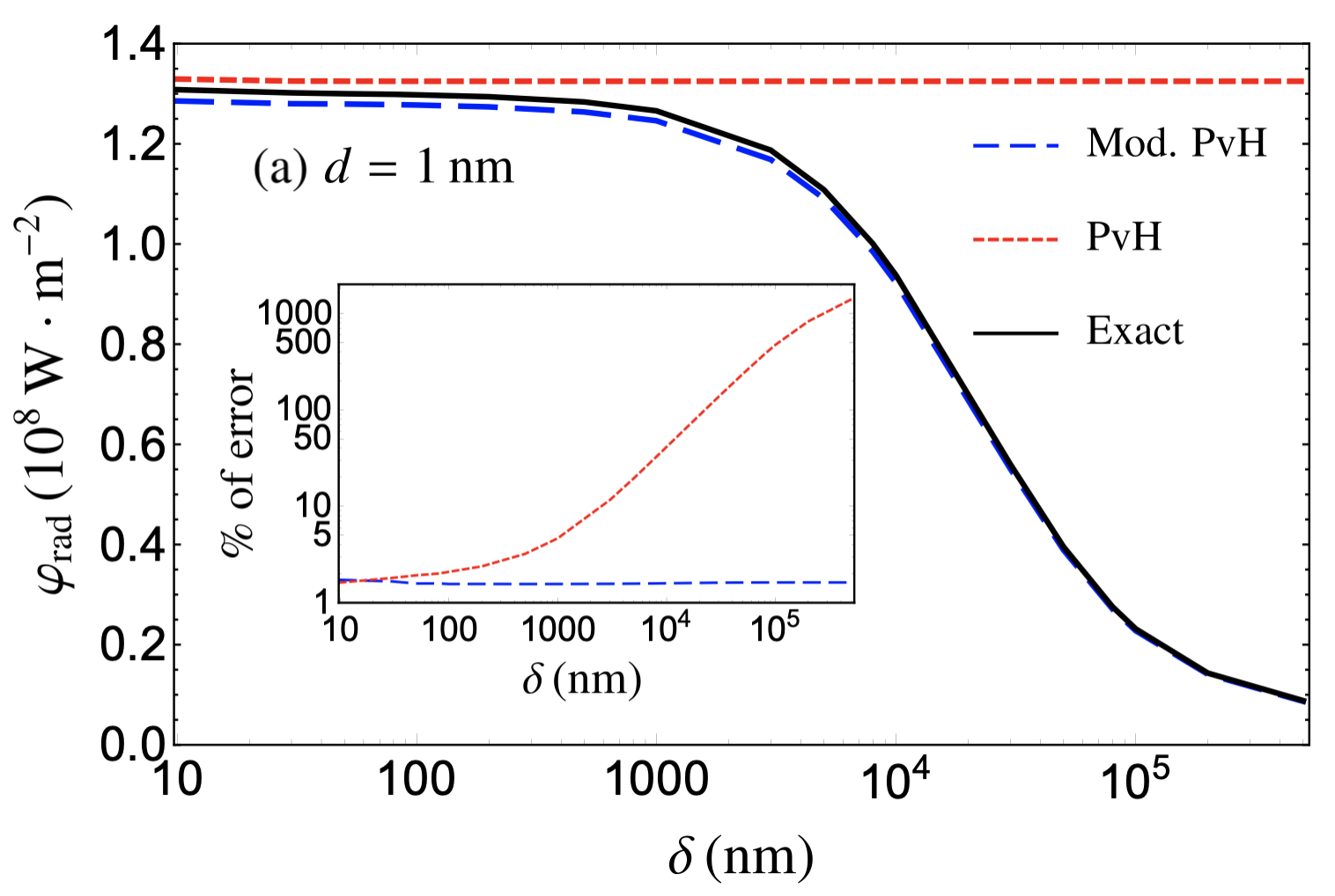}\hspace{1.5cm}
		\includegraphics[width=7cm]{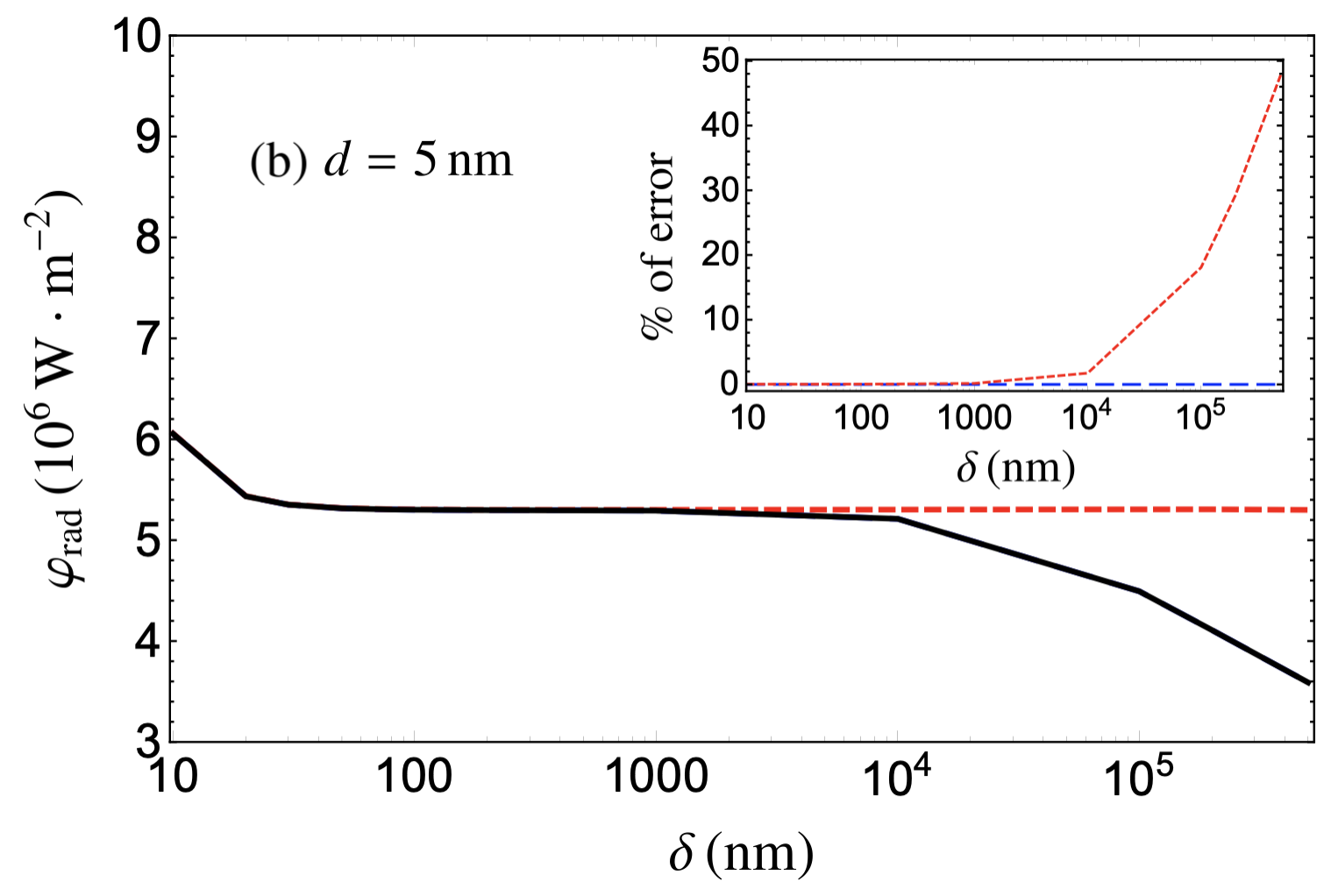}
	\caption{Radiative heat flux exchanged between two SiC slabs with respect to their thickness for a separation distance of (a) $d=1\,$nm and (b) $d=5\,$nm. We show the exact result (black line), the Polder and van Hove one (red dashed line, uniform temperatures $T_L=300\,$K and $T_R=400\,$K) and the modified Polder and van Hove flux (blue long-dashed line, uniform temperatures equal to the temperatures at the boundaries with the vacuum gap). Insets: absolute value of the error with respect to the PvH and modified PvH approaches. Reproduced from \cite{Reina2021a}.\label{fig_Boltzmann2}}
\end{figure}

In Fig.~\ref{fig_Boltzmann2} we address the impact of the temperature profiles on the exchange radiative flux. The exact result (solid black line) is compared to two approximate configurations: the Polder and van Hove result (i.e. conventional fluctuational electrodynamics ignoring the existence of a temperature profile, red dashed line) and to the \emph{modified Polder and van Hove} configuration (blue long-dashed line). The latter corresponds to a conventional fluctuational-electrodynamics configuration assuming that the temperature inside each body is uniform and equal to the temperature taken at the boundaries between it and vacuum.

These curves show that at the closest distance $d=1\,$nm the error in using standard fluctuational electrodynamics (ignoring conduction--radiation coupling) can be enormous, revealing once more the relevance of the coupling effect for large thicknesses and small distances. On the contrary, the modified approach reproduces relatively well the exact result. On one hand, this confirms that radiative heat transfer is mainly a surface effect, and thus almost entirely depends on the temperature at the boundaries between each body and vacuum. Nevertheless, albeit its fundamental interest, this has no direct practical use, since the knowledge of these boundary temperatures stems from and thus needs the solution of the full problem in the presence of coupling. Apart from being relevant for theory--experiment comparison, these coupling effects could be relevant in some applications, involging e.g. the thermalization of two bodies, as discussed in detail in \cite{Reina2021b}.

\section{The impact of geometry: the tip--plane configuration}\label{sec_geometry}

After investigating the role played by the thickness of the two bodies, it is interesting to address the impact of the geometrical configuration, not only to unveil possible fundamental issues but also because experiments are only rarely performed in the plane--plane configuration, which raises the serious experimental challenge of ensuring parallelism.

In a first work~\cite{Jin17b} we have studied conduction--radiation coupling between two nanorods. In this configuration we have highlighted, the appearance of a temperature profile, modifying in turn the exchanged radiative flux, along with a deviation from a linear temperature profile, even in the diffusive regime, due to the appearance of bulk polaritonic resonances, absent in the case of two parallel planes.

\begin{figure}
		\centering
		\includegraphics[width=9cm]{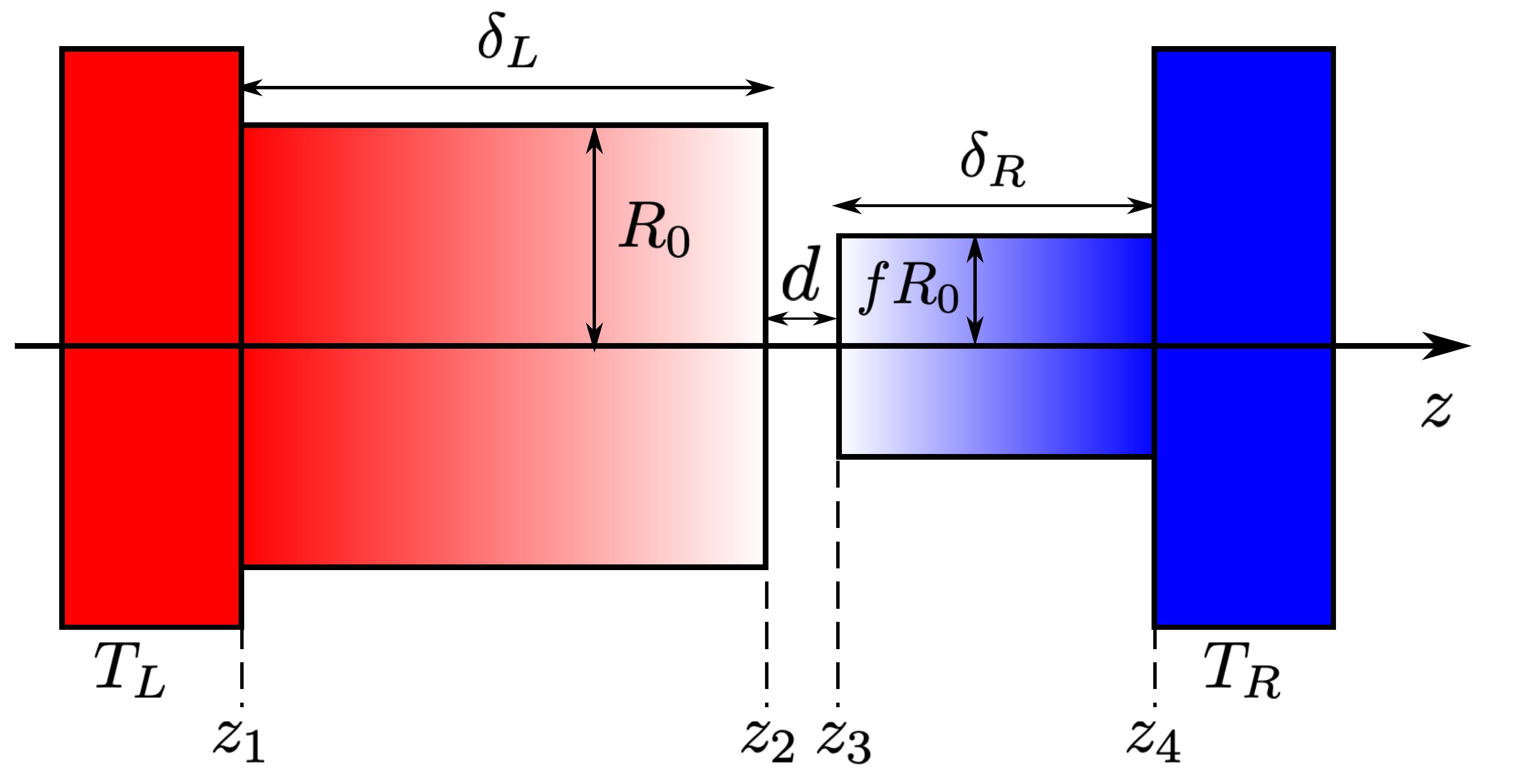}
	\caption{Scheme of the two-cylinder employed to simulate the tip--plane configuration. Reproduced from \cite{GharibAliBarura22}, with the permission of AIP Publishing.\label{fig_cylinder_geometry}}
\end{figure}

We focus here more in detail on a more recent work~\cite{GharibAliBarura22}, where we have analyzed the effect of conduction--radiation coupling in the tip--plane scenario, frequently employed in experiments. To this aim we have considered the geometrical configuration sketched in Fig.~\ref{fig_cylinder_geometry} where two cylinders of radius $R_0$ and $fR_0$ are placed in front of each other and separated by a vacuum gap of thickness $d$. The former (latter) has thickness $\delta_L$ ($\delta_R$) and is connected to a thermostat at temperature $T_L$ ($T_R$). The modulation of the factor $f$ allows to switch from the plane--plane configuration ($f=1$) to a tip--plane scenario, when $f$ tends to zero.

Inspired by the results discussed in Sec.~\ref{sec_Boltzmann}, we have considered large values for the two thicknesses ($\delta_L=\delta_R=100\,\mu$m) such that we are in the diffusive regime. We have then solved the heat equation in cylindrical coordinates, by assuming that the two cylinders do not exchange any flux across their lateral surfaces. Moreover, we have also assumed that radiative flux can be treated as a surface term (i.e. as a boundary condution) and, by following the Derjaguin approximation~\cite{Derjaguin68} (also known as proximity-force approximation), that twe two cylinders exchange heat radiatively only across the surface of the smaller cylinder. In other words, for cylinder 1 we have $\partial T(r,z_2)/\partial z=0$ for $fR_0<r<R_0$ while $\partial T(r,z_i)/\partial z=-\varphi(r)/\kappa$ for $i=2,3$, $\kappa$ being the thermal conductivity and $\varphi(r)$ the radiative flux locally exchanged at coordinate $r$. The solution can be obtained analytically under the furthe approximation that the exchanged flux can be written as
\begin{equation}\label{SimplFlux}
\varphi\simeq\frac{\gamma[T(0,z_2) - T(0,z_3)]}{d^2},
\end{equation}
i.e. it is uniform, it depends only on the two temperatures at the center of the two cylinder surfaces and on the separation distance as $d^{-2}$. This allows to obtain analytical expressionf for the temperature profiles in the cylinders
\begin{equation}\label{Texact}
		T(r, z) = \begin{cases}
			T_L - \frac{\gamma(T_L - T_R)}{\xi}\Biggl[f^2(z-z_1)+2 R_0f\sum_{k=1}^{\infty}\frac{J_1(f\alpha_k)}{\alpha_k^2 J_0^2(\alpha_k)}\frac{\sinh\Bigl[\frac{\alpha_k(z - z_1)}{R_0}\Bigr]}{\cosh\Bigl[\frac{\alpha_k(z_2 - z_1)}{R_0}\Bigr]}J_0\Bigl(\alpha_k\frac{r}{R}\Bigr)\Biggr], & z_1<z<z_2,\\
			T_R + \frac{\gamma(T_L - T_R)}{\xi} (z_4 - z), & z_3<z<z_4,\end{cases}
	\end{equation}
and for the exchanged radiative flux
\begin{equation}\label{eq:phid}
	\varphi(d,f,R_0) = \frac{\frac{\gamma(T_L - T_R)}{d^2}}{1+\frac{\gamma}{\kappa d^2}\Bigl[f^2\delta_L+\delta_R+2R_0f\Gamma\Bigl(f,\frac{\delta_L}{R_0}\Bigr)\Bigr]},
\end{equation}
$\delta_L=z_2-z_1$ ($\delta_R=z_4-z_3$) being the height of the larger (smaller) cylinder. In these expressions we have defined
\begin{equation}\begin{split}
		\xi&=\kappa d^2+\gamma(f^2\delta_L+\delta_R)+2\gamma R_0f\Gamma\Bigl(f,\frac{\delta_L}{R_0}\Bigr),\\
		\Gamma(f,\beta)& = \sum_{k=1}^{\infty}J_1(f\alpha_k)\tanh(\alpha_k\beta)/[\alpha_k^2 J_0^2(\alpha_k)].
	\end{split}
\end{equation}
As expected, Eq.~\eqref{eq:phid} allows to get back the results for two parallel slabs~\cite{Messina2016a} for $f=1$. As discussed more in detail in \cite{GharibAliBarura22}, going beyond the approximation described in Eq.~(\ref{SimplFlux}) allows to get numerical results in very good agreement, for the physical parameters taken into account below, with the analytical expressions.

\begin{figure}
		\centering
		\includegraphics[width=7cm]{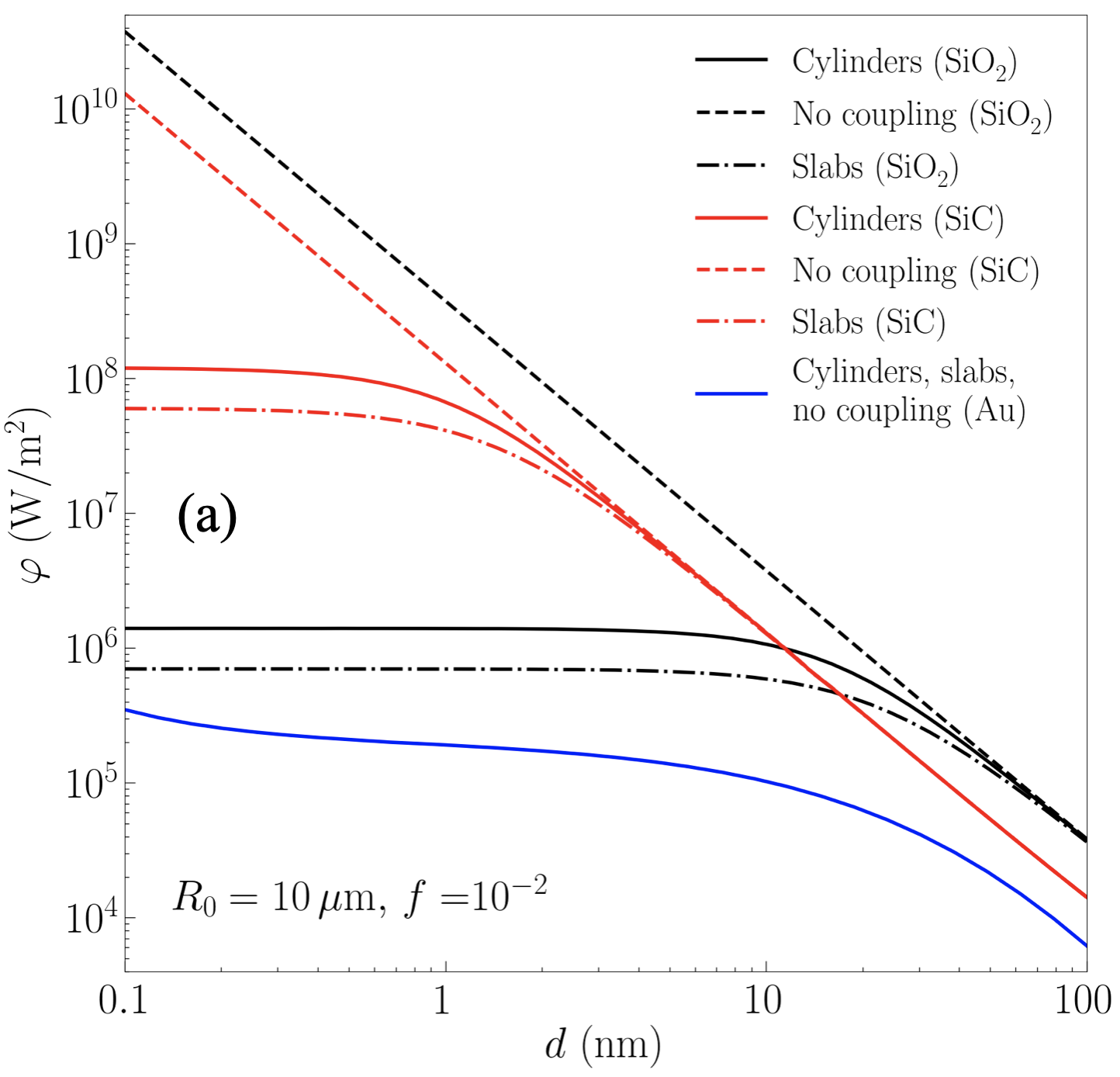}\hspace{1.5cm}
		\includegraphics[width=7.6cm]{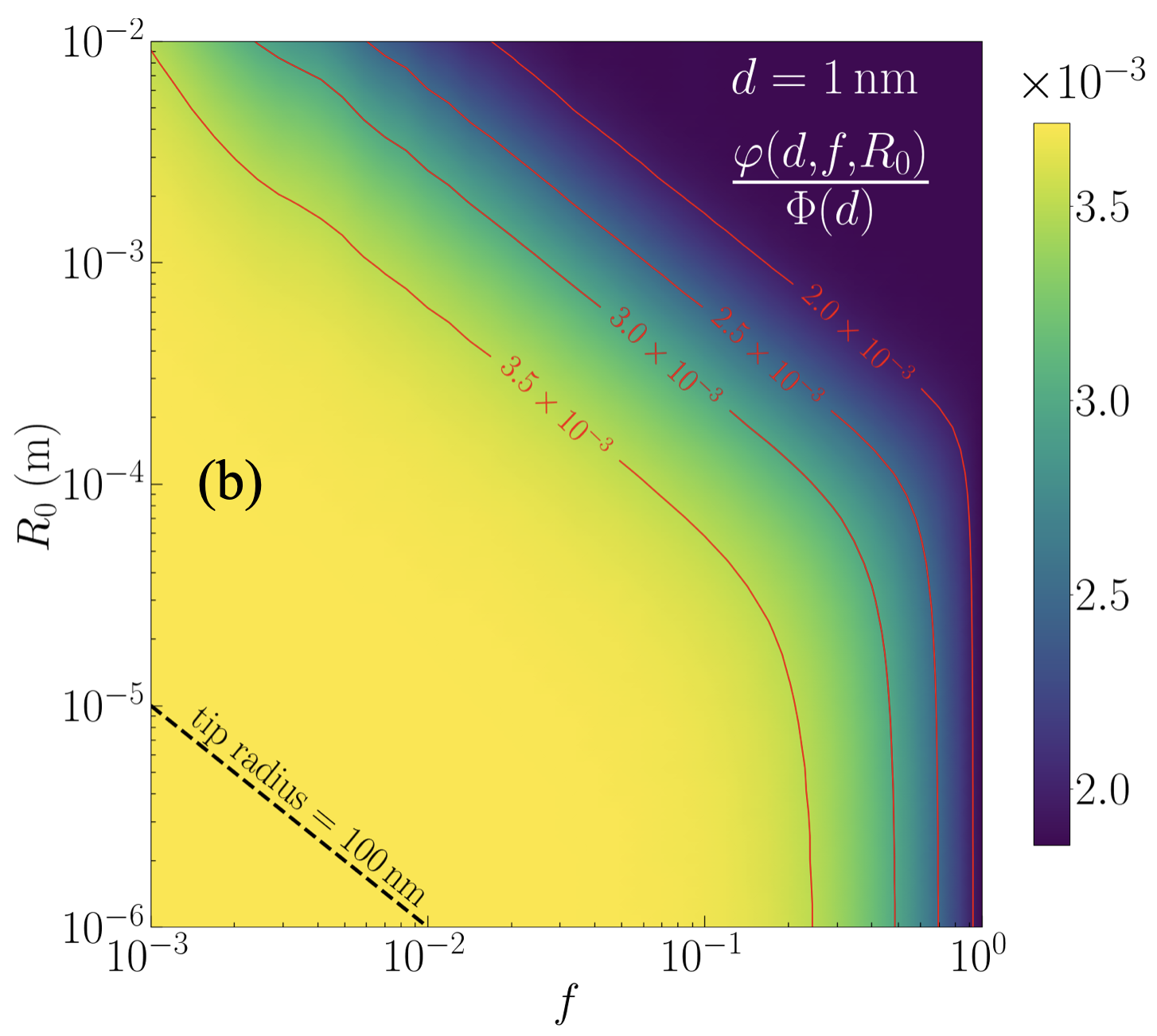}
	\caption{(a) Radiative heat flux as a function of $d$ in a tip\,-\,plane configuration (solid lines), slab-slab scenario (dot-dashed lines), and in the absence of coupling (dashed lines). The different colors correspond to different materials: black for SiO$_2$, red for SiC and blue for Au. (b) Ratio between fluxes in the presence and absence of coupling as a function of $R_0$ and $f$. The plot corresponds to two SiO$_2$ cylinders having $\delta_L=\delta_R=100\,\mu$m at distance $d=1\,$nm. Reproduced from \cite{GharibAliBarura22}, with the permission of AIP Publishing.\label{fig_cylinder}}
\end{figure}

The results for a configuration with $\delta_L=\delta_R=100\,\mu$m, $R_0=10\,\mu$m and $f=10^{-2}$ are shown in Fig.~\ref{fig_cylinder}(a), where they are compare to the configuration of absence of coupling and to the slab--slab scenario (corresponding to $f=1$).

More specifically, the figure shows the results for silica (SiO$_2$), silicon carbide (SiC) and gold bodies. In the latter scenario we observe that the curves in the absence and in the presence of coupling (both for slabs and cylinders) are indistinguishable. The reason behind this is that the radiative exchange in the case of gold is weak compared to the case of polar materials, mainly because the surface resonant modes (surface plasmons) supported by gold bodies lie in the ultraviolet region of the spectrum, and are thus not excited thermally around ambient temperature. As a result, no significant temperature profile and no impact on the radiative flux are expected.

On the contrary, based on what we already discussed in Sec.~\ref{sec_slabs}, for two polar materials radiative heat flux in the nanometer range of distances is supposed to compete with conduction. The results of Fig.~\ref{fig_cylinder}(a) show that the results for two cylinders qualitatively follow the ones for two slabs: the characteristic distance at which the curve in the presence of coupling deviates from the one in the absence of coupling is almost the same, while the value of the saturated ($d\to 0$) flux is only slightly higher than the one for two slabs. This allows to state that also in a tip--plane configuration the impact of conduction--radiation coupling is supposed to be observable, at least for polar materials, in the nanometer range of distances.

To conclude, we analyze in Fig.~\ref{fig_cylinder}(b) the combined effect of radius $R_0$ and radii fraction $f$. We observe that the ratio between flux in the presence [$\varphi(d,f,R_0)$] and in the absence [$\Phi(d)$] of coupling is of the order of $10^{-3}$ in a wide range of both parameters. Finally, the black dashed line in Fig.~\ref{fig_cylinder}(b) indicates that the flux correction is significant for all points defining a hypothetical tip radius of 100\,nm, corresponding to an experimentally reasonable value.

\section{Conclusions}\label{sec_conclusions}

We have reviewed our recent work on the coupling between conduction and radiation for two solids out of thermal equiibrium interacting in near-field regime. We have shown that, depending on the separation distance and the materials nature under scrutiny, this coupling can be at the origin of a non-negligible temperature profile inside each body (ignored in previous works), which can in turn induce a saturation of the radiative heat flux exchanged in the two bodies with respect to the predictions of conventional fluctuational electrodynamics. Because of the current possibility to explore experimentally distances in the nanometer range and below and of the ongoing miniaturization of a variety of technological devices, these results show that this coupling needs to be taken into account, both for the sake of theory--experiment comparison and in view of the design of innovative devices operating at the nanoscale.

\end{document}